\documentclass[
floatfix,
aps,
amsmath,
nofootinbib,
onecolumn,
10pt,
]{revtex4}

\usepackage{graphicx}
\usepackage{bm}
\usepackage{rotating}
\usepackage{array}

\begin{document}

\title{Numerical Relativity Investigation of the Effects of Gravitational
    Waves on the Inhomogeneity of the Universe}

\author{
Ke Wang \footnote{wangke@itp.ac.cn}
}

\affiliation{
National Astronomical Observatories, \\Chinese Academy of Sciences,  20A Datun Road, Beijing 100012, China\\
}

\date{\today}

\begin{abstract}
We numerically integrate the Einstein's equations for a spatially flat Friedmann-Lemaire-Robertson-Walker (FLRW) background spacetime with a 
spatial curvature perturbation and evolving primordial tensor perturbations using the Einstein Toolkit. We find that although the primordial tensor perturbation doesn't play an important role in the evolution of the overdensity produced by the scalar perturbation, there is an obvious imprint left by the primordial tensor perturbation on the distribution of the fractional density perturbation in the nonlinear region. This imprint may be a possible probe of a gravitational waves background in the future.

\end{abstract}


\maketitle


\section{Introduction}

Inflation \cite{Starobinsky:1980te,Guth:1980zm} predicts that there is a stochastic gravitational waves (GW) background. Therefore, it's possible to test inflation scenario experimentally through detection of such a GW background. So far, the B-mode polarization of the cosmic microwave background (CMB) is the most promising probe of this GW background \cite{Seljak:1996gy,Kamionkowski:1996zd,Kamionkowski:2015yta}. In the future, 21cm HI emission from the dark ages will be a complementary and even more sensitive probe of this GW background \cite{Book:2011dz,Masui:2010cz}. Furthermore, there are some not very competitive probes of this GW background, including weak lensing shear \cite{Dodelson:2003bv,Dodelson:2010qu} and other large-scale structure observables \cite{Jeong:2012nu,Schmidt:2012nw}. The goal of this paper is to study the effect of gravitational waves on the inhomogeneity of the Universe with numerical relativity, thereby proposing a possible probe of a GW background.

Although cosmological principle points out that the Universe is homogeneous and isotropic on large scales and can be described by a Friedmann-Lemaitre-Robertson-Walker (FLRW) model, it is inhomogeneous and anisotropic on scales smaller than $\sim80h^{-1}\textrm{Mpc}$ \cite{Yadav:2010cc,Scrimgeour:2012wt} today. These inhomogeneities can induce nonlinear general relativistic effects which may be detected by forthcoming cosmological surveys \cite{Amendola:2016saw,Maartens:2015mra,Ivezic:2008fe}. Moreover, these nonlinear general relativistic effects on small scales may be accompanied by unexpected nonperturbative behavior on larger scales. This ``backreaction" is believed to be the reason of the recent cosmic expansion by \cite{Buchert:1999er,Kolb:2004am,Buchert:2007ik,Rasanen:2011ki,Clarkson:2011zq,Buchert:2011sx,Buchert:2015iva,Green:2015bma}. We will, however, use a restricted simulation setup by using a fixed FLRW background and subjecting perturbations to periodic boundary conditions, which suppresses the large-scale backreaction effect on the global evolution of the background spacetime \cite{Buchert:2017obp}.

Usually, the linear perturbation theory of General Relativity (GR) is used on large scales and Newtonian N-body simulations (or Newtonian gravity) provide a very good approximation on small scales. To study the nonlinear general relativistic effects, one can do general relativistic N-body simulations as \cite{Adamek:2013wja}. Undoubtedly, the direct numerical integration of Einstein's equation is the only way without any systematic errors and approximation to study the Universe on all scales. The first cosmological work that is fully non-linear, fully relativistic and does not impose symmetries or dimensional-reductions has been done by \cite{Giblin:2015vwq,Mertens:2015ttp} using \textsf{C\textsc{osmo}GR\textsc{a}PH}. For cosmological purpose, soon afterward, \cite{Bentivegna:2015flc} turned to the wide-used \textsf{Einstein Toolkit} \cite{Loffler:2011ay} to integrate Einstein's equation. \cite{Macpherson:2016ict} also studied the inhomogeneous cosmology with \textsf{Einstein Toolkit} by developing a new thorn, \textsf{FLRWSolver}.

Here, our work is based on: a new thorn \textsf{CFLRWSolver} (a \textsf{C} language counterpart of \textsf{FLRWSolver}) which takes the tensor perturbation into consideration and initializes an almost FLRW Universe with saclar and tensor perturbations; a self-developing thorn \textsf{CFLRWAnalysis} which calculates several derived varivables including the comoving time, scale factor, the tensor component of perturbation and the distribution of overdensity at the end of each evolution step; the thorn \textsf{McLachlan} \cite{Brown:2008sb,Reisswig:2010cd,code} which evolves spacetime using the Baumgarte-Shapiro-Shibata-Nakamura (BSSN) formalism \cite{Baumgarte:1998te,Shibata:1995we,Alcubierre:2000xu} and the thorn \textsf{GRHydro} which evolves the hydrodynamical system \cite{Moesta:2013dna,Baiotti:2004wn,Hawke:2005zw}.

This paper is organized as follows. In Sec.~\ref{pert}, we give the evolution equations of the FLRW background spacetime and scalar and tensor perturbations through solving the zero-order and first-order of Einstein equations and solutions to them. In Sec.~\ref{initial}, we give the initial conditions of the system with small perturbations needed by the thorn \textsf{CFLRWSolver}. In Sec.~\ref{results}, we analyse the results of simulations provided by the thorn \textsf{CFLRWAnalysis}. At last, a brief summary and discussion are included in Sec.~\ref{summary}.

In this paper, we adopt the following conventions: Greek indices run in \{0, 1, 2, 3\}, Latin indices run in \{1, 2, 3\} and repeated indices implies summation and we are in a geometric unit system with $G=c=1$.
\section{Cosmological Perturbations}
\label{pert}

For a spatially flat FLRW background spacetime, the line element is
\begin{equation}
ds^2=a^2(\eta)[-d\eta^2+\delta_{ij}dx^idx^j],
\end{equation}
where $\eta$ is the conformal time, $a$ is the scale factor and $\delta_{ij}$ is the identity matrix.
In the conformal Newtonian gauge, the line element that includes both the scalar and tensor perturbations to the metric is
\begin{equation}
\label{metric}
ds^2=a^2(\eta)[-(1+2\Psi)d\eta^2+(1-2\Phi)\delta_{ij}dx^idx^j+h_{ij}dx^idx^j],
\end{equation}
where $\Psi$ is the Newtonian potential, $\Phi$ the spatial curvature perturbation and $h_{ij}$ is a divergenceless, traceless and symmetric tensor.
And for a perfect fluid without the anisotropic stress tensor, its energy-momentum tensor with density $\rho=\rho_0+\rho_1$, isotropic pressure $P=P_0+P_1$ and 4-velocity $u^\mu=a^{-1}[1-\Psi,v^1_1,v^2_1,v^3_1]$ is
\begin{equation}
\label{EMtensor}
T_{\mu\nu}=(\rho+P)u_\mu u_\nu+Pg_{\mu\nu}.
\end{equation}
The Einstein equations relate the spacetime curvature to the energy-momentum tensor as
\begin{equation}
G_{\mu\nu}=8\pi T_{\mu\nu}.
\end{equation}

The zero-order Einstein equations give the Friedmann constraint and evolution equation for the FLRW background spacetime
\begin{eqnarray}
\label{back}
\mathcal{H}^2&=&\frac{8\pi}{3}a^2\rho_0,\\ \nonumber
\mathcal{H}'&=&-\frac{4\pi}{3}a^2(\rho_0+3P_0),
\end{eqnarray}
where a prime represents a derivative with respect to the conformal time. According to \cite{Macpherson:2016ict}, the dust ($P\ll\rho$) solution to (\ref{back}) is
\begin{eqnarray}
\label{sovleback}
a&=&a_{\mathrm{init}}\xi^2,\\\nonumber
\rho_0&=&\rho_{0,\mathrm{init}}\xi^{-6},\\\nonumber
\xi&=&1+\sqrt{\frac{2\pi\rho_0*}{3a_{\mathrm{init}}}}\eta,
\end{eqnarray}
where $a_{\mathrm{init}}$ and $\rho_{0,\mathrm{init}}$ are the values of $a$ and $\rho_0$ at $\eta=0$ respectively, $\xi$ is the scaled conformal time and $\rho_0*=\rho_0 a^3$ is the conserved comoving density.

From the first-order perturbed Einstein equations, we derive equations describing scalar metric perturbations as \cite{Macpherson:2016ict,Malik:2008im}
\begin{eqnarray}
\label{scalar}
\nabla^2\Phi-3\mathcal{H}(\Phi'+\mathcal{H}\Psi)&=&4\pi a^2 \rho_1,\\\nonumber
\mathcal{H}\partial_i\Psi+\partial_i\Phi'&=&-4\pi a^2(\rho_0+P_0)\delta_{ij}v^j_1,\\
\nonumber
\Phi&=&\Psi,\\
\nonumber
\Phi''+3\mathcal{H}\Phi'+(2\mathcal{H}'+\mathcal{H}^2)\Phi&=&4\pi a^2P_1.
\end{eqnarray}
\cite{Macpherson:2016ict} also gives the dust ($P\ll\rho$) solution to (\ref{scalar}) for the growing mode as
\begin{eqnarray}
\label{sovlescalar}
\Phi&=&f(x^i),\\\nonumber
\frac{\rho_1}{\rho_0}&=&C_1\xi^2\nabla^2f(x^i)-2f(x^i),\\\nonumber
v_1^i&=&C_2\xi\partial^if(x^i),
\end{eqnarray}
where $f(x^i)$ is an arbitrary function of space, $C_1=\frac{a_{\mathrm{init}}}{4\pi\rho_0*}$ and $C_2=-\sqrt{\frac{a_{\mathrm{init}}}{6\pi\rho_0*}}$.

From the spatial part of the first-order perturbed Einstein equations, we have a wave equation
\begin{equation}
h_{ij}''+2\mathcal{H}h_{ij}'-\nabla^2h_{ij}=0.
\end{equation}
One can expand the tensor perturbation in plane waves
\begin{equation}
h_{ij}(\vec{x},\eta)=\int \frac{d^3k}{(2\pi)^3}h^{s}_k(\eta)\varepsilon^s_{ij}e^{i\vec{k}\cdot\vec{x}},
\end{equation}
where $\varepsilon^s_{ij}$ with $s=\times,+$ are transverse and traceless polarization tensors and each of $h^{s}_k(\eta)$ evolves independently and satisfies
\begin{equation}
{h^s_k}''+2\mathcal{H}{h^s_k}'+k^2h^s_k=0.
\end{equation}
According to \cite{Wang:1995kb}, for modes inside the horizon during matter dominated era, the exact solution is
\begin{equation}
\label{sovletensor}
h_{k}^s(\eta+\eta_0)=3h_{k}^s(0)\frac{\sin[k(\eta+\eta_0)]-[k(\eta+\eta_0)]\cos[k(\eta+\eta_0)]}{[k(\eta+\eta_0)]^3},
\end{equation}including $a$ ,$\rho_0$ and $\eta$
where $\eta+\eta_0$ is the comoving size of horizon.
\section{Initial Conditions}
\label{initial}

To integrate Einstein equations, \textsf{Einstein Toolkit} turns to the metric in the form of $(3+1)$ formalism
\begin{equation}
ds^2=-\alpha^2dt^2+\gamma_{ij}(dx^i+\beta^idt)(dx^j+\beta^jdt),
\end{equation}
where $\alpha$ is the lapse function, $\beta^i$ is the shift vector and $\gamma_{ij}$ is the spatial metric and evolves depending on the extrinsic curvature $K_{ij}$ as
\begin{equation}
(\partial_t-\mathcal{L}_{\vec{\beta}})\gamma_{ij}=-2\alpha K_{ij}.
\end{equation}

Now we will use \textsf{CFLRWSolver} to initialize an almost FLRW Universe with small perturbations. First, we should relate the groups of gird function for basic spacetime variables in the thorn \textsf{ADMBase} to the variables in (\ref{metric}):
\begin{eqnarray}
\gamma_{ij}&=&a^2[(1-2\Phi)\delta_{ij}+h_{ij}],\\\nonumber
K_{ij}&=&\frac{2a'[(1-2\Phi)\delta_{ij}+h_{ij}]-2a\Phi'\delta_{ij}+ah_{ij}'}{-2\sqrt{1+2\Psi}}.\\\nonumber
\end{eqnarray}including $a$ ,$\rho_0$ and $\eta$
where we have set $dt=\frac{a\sqrt{1+2\Psi}}{\alpha} d\eta$ and $\beta^i=0$.  Here we choose the harmonic slicing
\begin{equation}
\label{gauge}
\partial_t\alpha=-\frac{1}{4}\alpha^2K,
\end{equation}
which describes the evolution of $\alpha$. Since $K$ is negative in our simulation, this choice of foliation will be with high computational efficiency.
Then, we relate the basic variables and grid functions for hydrodynamics evolutions in the thorn \textsf{HydroBase} to the variables in (\ref{EMtensor}):
\begin{eqnarray}
\rho&=&\rho_0+\rho_1,\\\nonumber
P&=&P_0+P_1,\\\nonumber
v^i&=&\frac{v_1^i}{a},\\\nonumber
\Gamma&=&\left(1-\gamma_{ij}\frac{v_1^i}{a}\frac{v_1^j}{a}\right)^{-\frac{1}{2}}.
\end{eqnarray}
According to (\ref{sovleback}), (\ref{sovlescalar}) and (\ref{sovletensor}), we set the initial conditions for a dust system with periodic boundary conditions at $t=\eta=0$ as
\begin{eqnarray}
\Phi &=&\Phi_0\sum_{i=1}^3\sin\left(\frac{2\pi x^i}{l}\right),
\\\nonumber
\alpha&=&\sqrt{1+2\Phi},
\\\nonumber
h_{ij}&=&3h_{\frac{2\pi}{L}}^{s}(0)\frac{L^3\sin(\frac{2\pi\eta_0}{L})-2\pi L^2\eta_0\cos(\frac{2\pi\eta_0}{L})}{(2\pi\eta_0)^3}\cos\left[\frac{2\pi (z+125)}{L}\right]\varepsilon^{s}_{ij},
\\\nonumber
h_{ij}'&=&3h_{\frac{2\pi}{L}}^{s}(0)\frac{[(2\pi\eta_0)^4L^2-3(2\pi\eta_0)^2L^4]\sin(\frac{2\pi\eta_0}{L})+3(2\pi\eta_0L)^3\cos(\frac{2\pi\eta_0}{L})}{(2\pi\eta_0)^6}\frac{2\pi}{L}\cos\left[\frac{2\pi (z+125)}{L}\right]\varepsilon^{s}_{ij},
\\\nonumber
\gamma_{ij}&=&a_{\mathrm{init}}^2(\delta_{ij}-2\Phi\delta_{ij}+h_{ij}),
\\\nonumber
K_{ij}&=&\frac{\sqrt{\frac{32\pi\rho_{0,\mathrm{init}}}{3}}a_{\mathrm{init}}^2(\delta_{ij}-2\Phi\delta_{ij}+h_{ij})+a_{\mathrm{init}}h_{ij}'}{-2\alpha},
\\\nonumber
\rho&=&\rho_{0,\mathrm{init}}-\rho_{0,\mathrm{init}}\left[\left(\frac{2\pi}{l}\right)^2C_1+2\right]\Phi,
\\\nonumber
v^i&=&\frac{1}{a_{\mathrm{init}}}\frac{2\pi}{l}C_2\Phi_0\cos\left(\frac{2\pi x^i}{l}\right),
\\\nonumber
\Gamma&=&\left(1-\gamma_{ij}v^iv^j\right)^{-\frac{1}{2}},
\end{eqnarray}
where $l=500$ is the half length of one side of our simulation box with $x^i$ in $[-375, 625]$, $\Phi_0$ is the amplitude of spatial curvature perturbation and $h_{\frac{2\pi}{L}}^{s}(0)$ is the amplitude of monochromatic primordial tensor perturbation with wave number $k=\frac{2\pi}{L}$ before horizon-crossing. For simplicity, we set the initial scale factor $a_{\mathrm{init}}=1$. And we set $L<4000$ and $\rho_{0,\mathrm{init}}=10^{-6}$ so that $\frac{2\pi\eta_0}{L}>1$ which implies the tensor perturbation has crossed inside the horizon $\eta_0\simeq2\sqrt{\frac{3}{8\pi\rho_{0,\mathrm{init}}}}$ at the beginning of simulation. To keep the linear approximation remains valid and save the computational time as much as possible, we set $\Phi_0=10^{-4}$ which means the density contrast is about $10^{-3}$. Here we will run three simulations at resolution $100^3$ with $h_{\frac{2\pi}{L}}^{s}(0)=0$, $h_{\frac{2\pi}{500}}^{+}(0)=10^{-3}$ and $h_{\frac{2\pi}{500}}^{\times}(0)=10^{-3}$ to study the effect of tensor perturbation on the inhomogeneity of universe.

\section{Results}
\label{results}
We can analyse the effect of tensor perturbation by comparing the outputs of several variables in the \textsf{CFLRWAnalysis} derived from the basic ones in the \textsf{ADMBase} and \textsf{HydroBase} from those simulations performed above directly. Firstly, the relation $dt=\frac{a\sqrt{1+2\Psi}}{\alpha} d\eta$ gives
\begin{equation}
\xi(t)=\left(\sqrt{\frac{6\pi\rho_{0,\mathrm{init}}}{1+2\Phi}}\int\alpha(t)dt+1\right)^{1/3},
\end{equation}
from which we can use (\ref{sovleback}) to obtain the evolution of $a$ and $\rho_0$ with the comoving time $\eta$ , hence the evolution of $h^{\times}(\eta_0+\eta)$
\begin{equation}
h^{\times}(\eta_0+\eta)=\frac{\gamma_{12}(\eta)}{a^2(\eta)},
\end{equation}
and $\frac{\rho_1}{\rho_0}(\eta)$
\begin{equation}
\frac{\rho_1}{\rho_0}(\eta)=\frac{(\rho_1+\rho_0)(\eta)-\rho_0(\eta)}{\rho_0(\eta)},
\end{equation}
where $\gamma_{12}(t)$ and $(\rho_1+\rho_0)(t)$ are basic variables in \textsf{ADMBase} and \textsf{HydroBase}. Since the background quantities including $a$, $\rho_0$ and $\eta$ are slightly space-dependent in an inhomogeneous spacetime, especially in the nonlinear region in a dynamical gauge like (\ref{gauge}), here these quantities are the average ones taken across the simulation box, which will introduce an element of inaccuracy and gauge dependence.

\begin{figure}[]
\begin{center}
\includegraphics[scale=0.5]{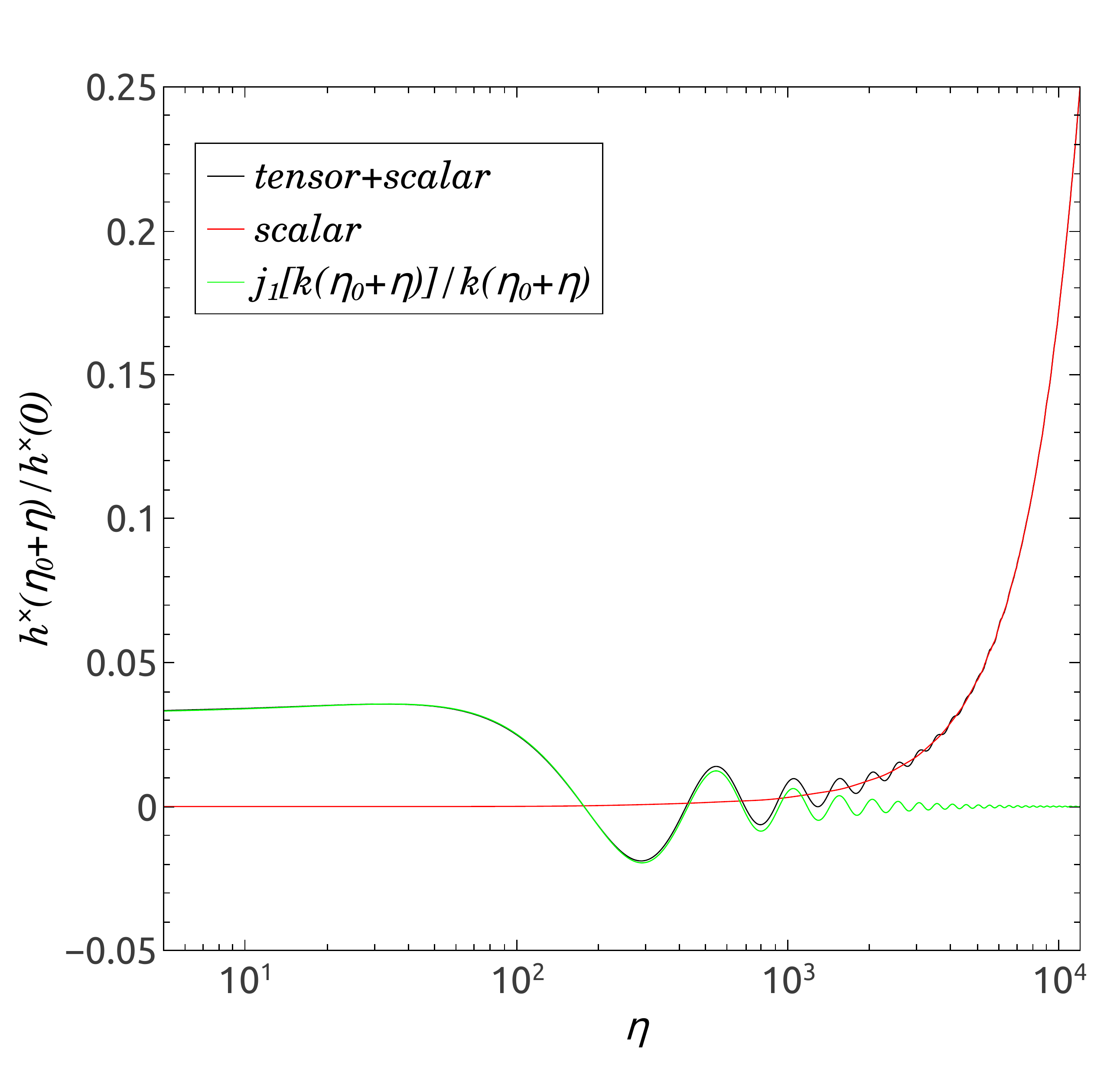}
\end{center}
\caption{The evolution of $\frac{h^{\times}(\eta_0+\eta)}{h^{\times}(0)}$ at the point $(0,0,375)$ of our simulation box (black and red solid curves).}
\label{fig:h}
\end{figure}

Fig.~\ref{fig:h} shows the evolution of $\frac{h^{\times}(\eta_0+\eta)}{h^{\times}(0)}$ at the point $(0,0,375)$ \footnote{Since we set the initial conditions of $\frac{\rho_1}{\rho_0}$ and $h^{\times}$ as $-\sin(\frac{2\pi x^i}{500})$-like and $\cos[\frac{2\pi (z+125)}{500}]$-like respectively, here we choose the particular locations for extracting the results where the maximums are obtained.} of our cubic domain in two simulations with $h_{\frac{2\pi}{L}}^{s}(0)=0$ (red solid curve) and $h_{\frac{2\pi}{500}}^{\times}(0)=10^{-3}$ (black solid curve). We can see that the scalar perturbations can produce the tensor perturbations due to the nonlinear effects (red solid curve) as pointed by \cite{Baumann:2007zm}, and the evolution of primordial tensor perturbation (black solid curve) follows the exact solution $\frac{j_1[k(\eta_0+\eta)]}{k(\eta_0+\eta)}$ (green solid curve), where $j_1(z)=\frac{\sin z-z\cos z}{z^2}$ is the spherical Bessel functions of order one.

\begin{figure}[]
\begin{center}
\includegraphics[scale=0.3]{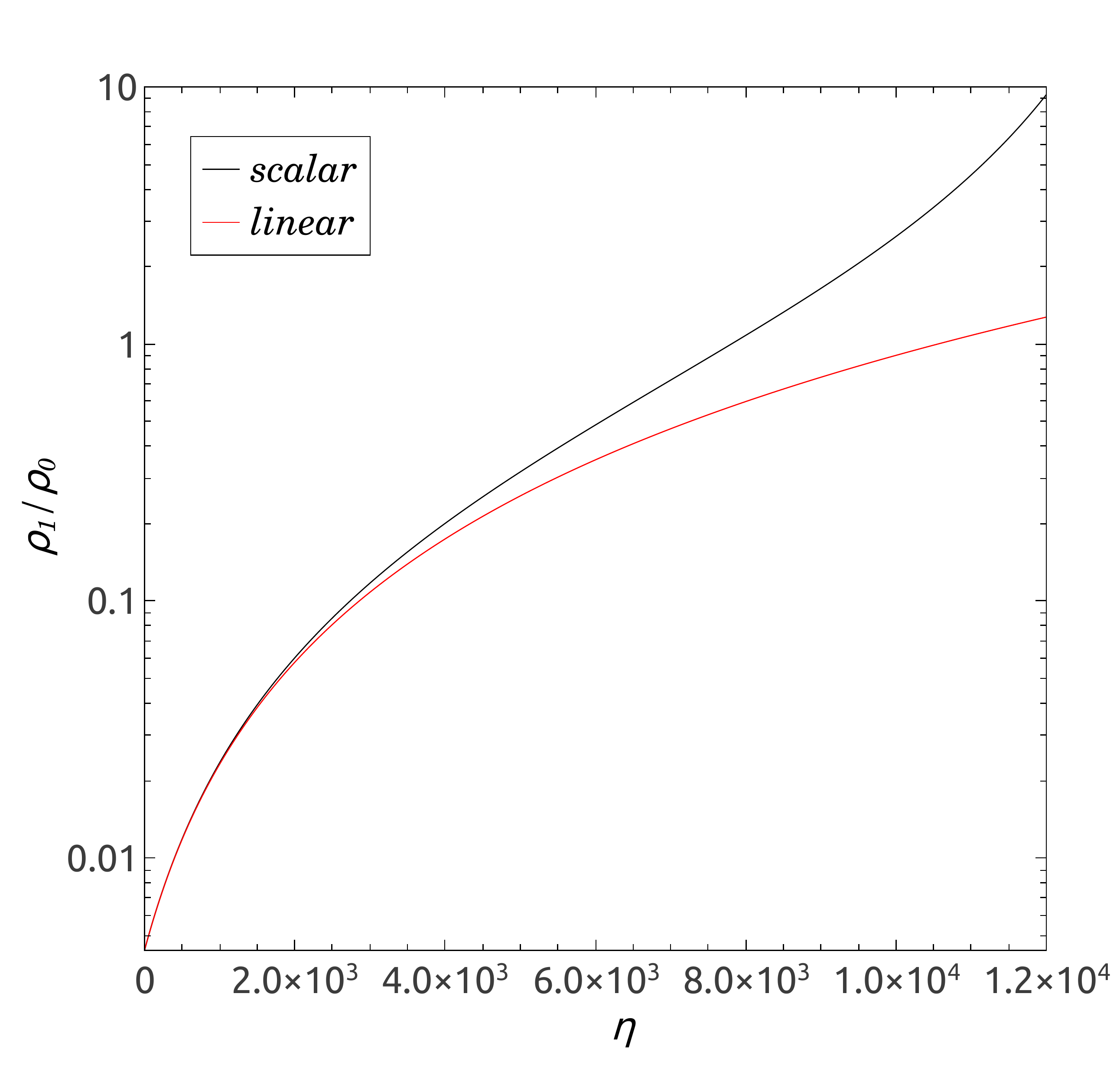}
\includegraphics[scale=0.3]{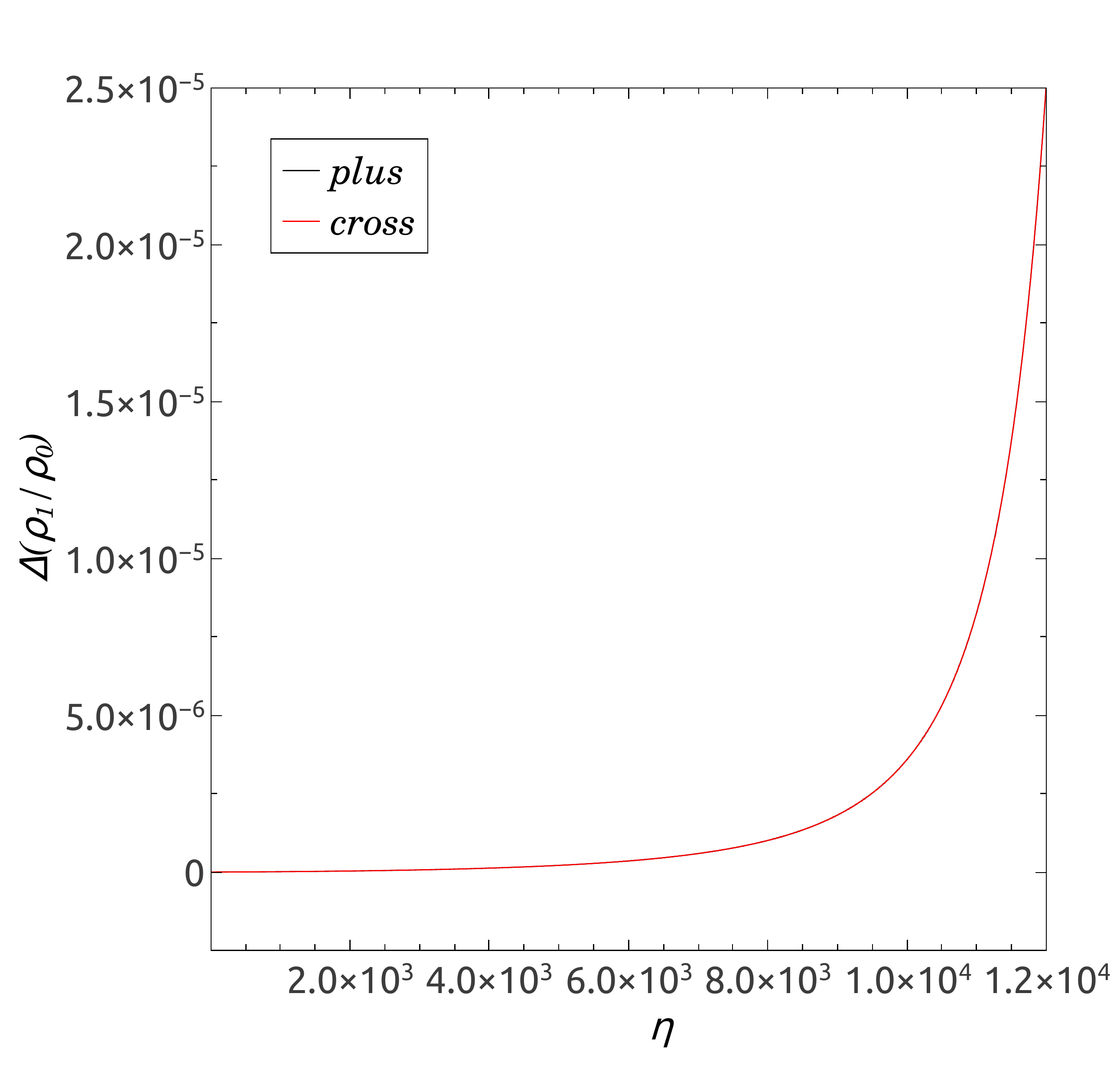}
\end{center}
\caption{The evolution of $\frac{\rho_1}{\rho_0}$ at the point $(375,375,375)$ of our simulation box without primordial tensor perturbations (left), and the primordial tensor perturbation's contribution to $\frac{\rho_1}{\rho_0}$ at the point $(375,375,375)$ (right).}
\label{fig:overdense}
\end{figure}

Fig.~\ref{fig:overdense} shows the evolution of $\frac{\rho_1}{\rho_0}$ at the point $(375,375,375)$ of our simulation box without primordial tensor perturbations (left), and the primordial tensor perturbation's contribution to $\frac{\rho_1}{\rho_0}$ at the point $(375,375,375)$ (right). We can see that the simulation result (black solid curve) deviates from the linear analytic solution (red solid curve) due to the nonlinear effects. Moreover, even though the primordial tensor perturbation die off quickly after the horizon-crossing as shown in Fig.~\ref{fig:h}, their contribution to $\frac{\rho_1}{\rho_0}$ grow up quickly in nonlinear regions.

\begin{figure}[]
\begin{center}
\includegraphics[scale=0.3]{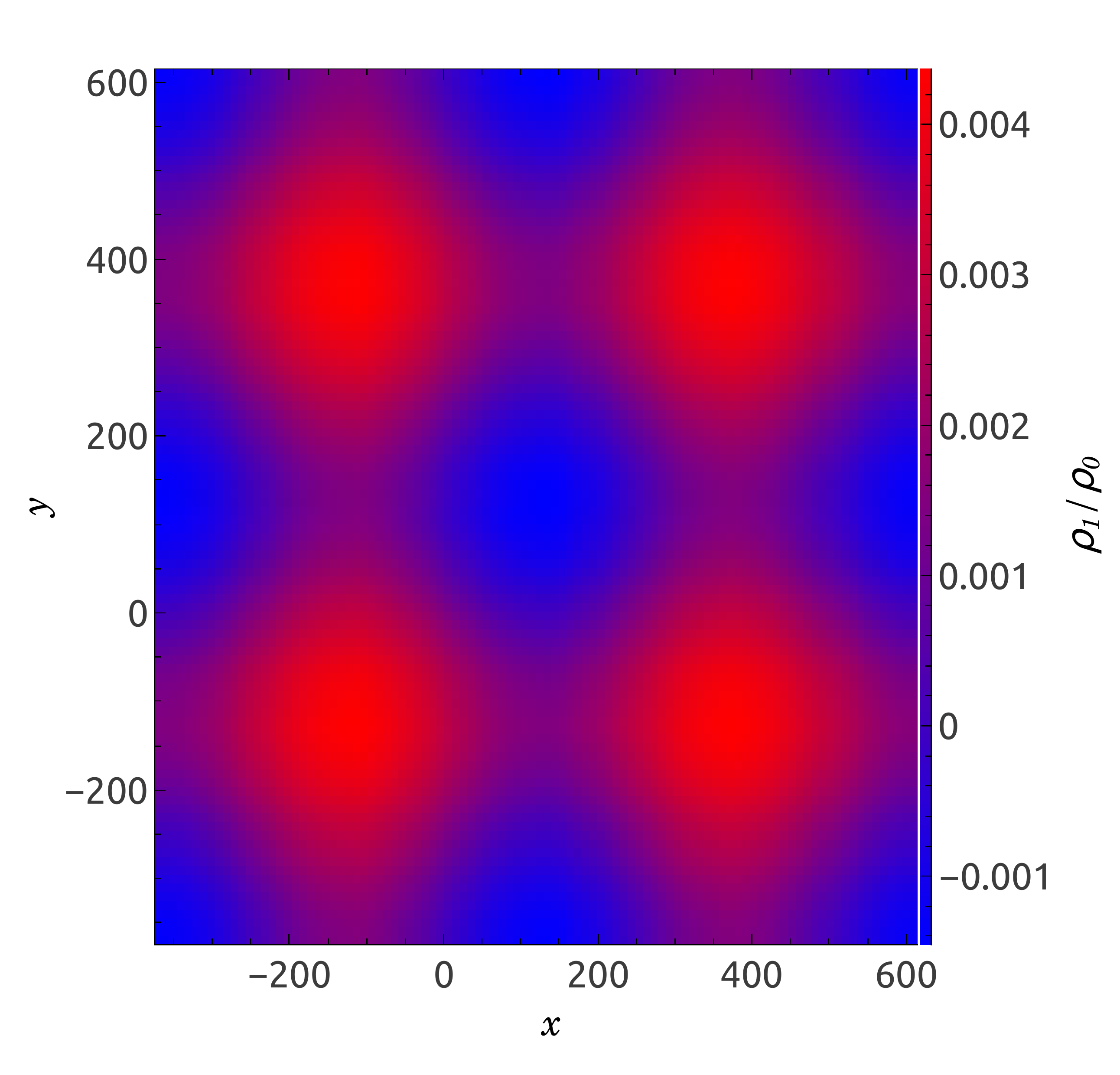}
\includegraphics[scale=0.3]{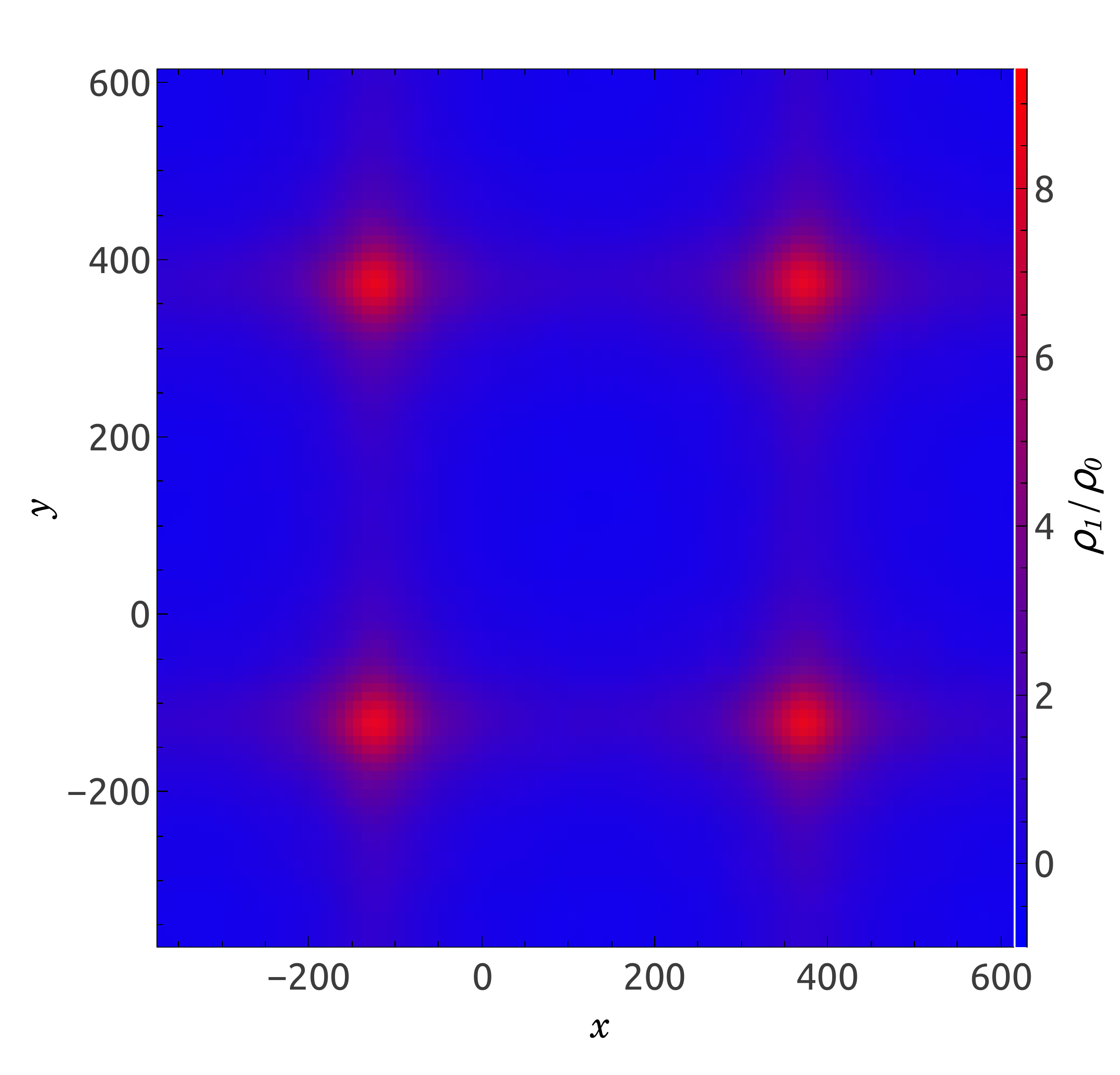}
\end{center}
\caption{The distribution of $\frac{\rho_1}{\rho_0}$ on the x-y plane of $z=375$ at the beginning and end of simulation without primordial tensor perturbations.}
\label{fig:xy}
\end{figure}

\begin{figure}[]
\begin{center}
\includegraphics[scale=0.3]{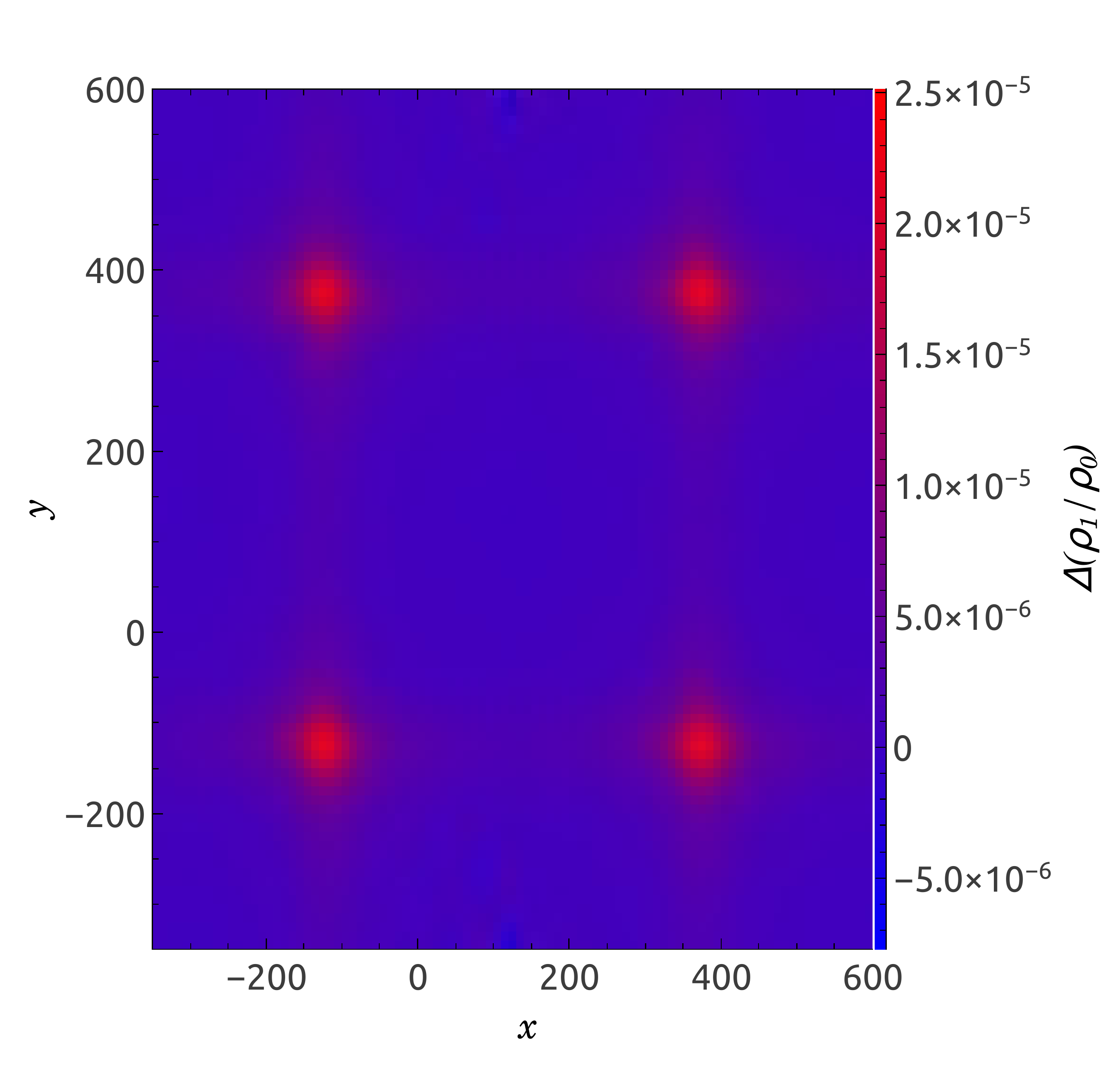}
\includegraphics[scale=0.3]{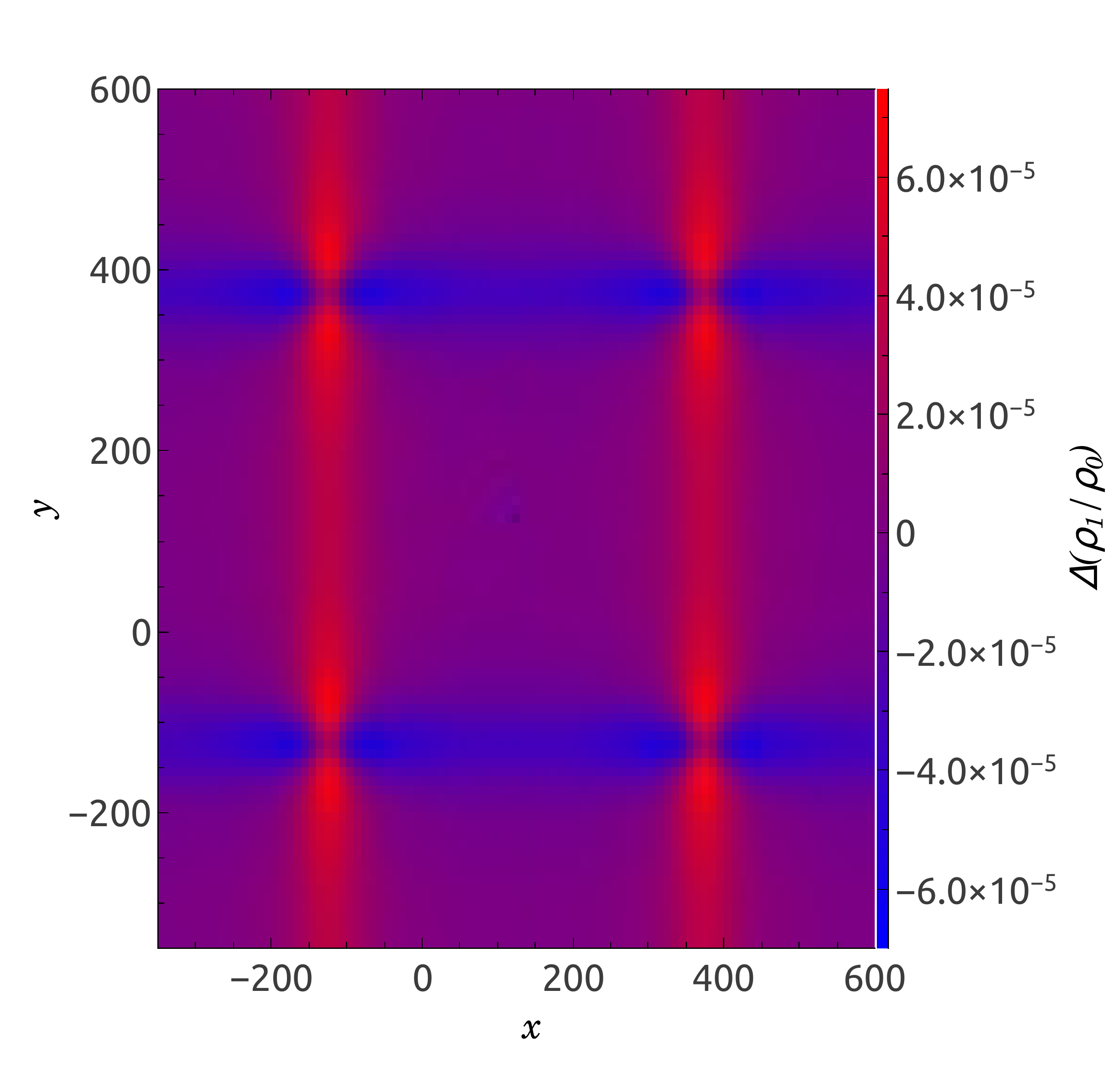}
\end{center}
\caption{The final contribution of primordial tensor perturbation with just $h^{\times}$ component (left) or $h^{+}$ component (right) to $\frac{\rho_1}{\rho_0}$.}
\label{fig:xy1}
\end{figure}

Fig.~\ref{fig:xy} shows the distribution of $\frac{\rho_1}{\rho_0}$ on the x-y plane of $z=375$ at the beginning $(\eta=0)$ and end $(\eta=1200)$ of simulation without primordial tensor perturbations. We can see that the locations of the maximum of $\frac{\rho_1}{\rho_0}$ are fixed at points $(-125,-125,375)$, $(375,-125,375)$, $(375,375,375)$ and $(-125,375,375)$. That is to say there is almost no interaction between the perturbation peaks.
The left one of Fig.~\ref{fig:xy1} shows the contribution of primordial tensor perturbation with just $h^{\times}$ component to $\frac{\rho_1}{\rho_0}$, while the right one shows the contribution from primordial tensor perturbation with only $h^{+}$ at the end of our simulation. We can see that the contributions of both cases are too small to modify the right one of Fig.~\ref{fig:xy}. However, primordial tensor perturbation with different component do leave a characteristic imprint on the distribution of $\frac{\rho_1}{\rho_0}$: the tensor perturbation with $h^{\times}$ enhances the overdensity; the tensor perturbation with $h^{+}$ enhances the overdensity alone the lines $(-125,y,375)$ and $(375,y,375)$ but suppresses the overdensity alone the lines $(x,-125,375)$ and $(x,375,375)$.

\section{Summary and discussion}
\label{summary}

We have performed three simulations using Einstein Toolkit in this paper: The first one gives the evolution of overdensity $\frac{\rho_1}{\rho_0}$ in a spatially flat FLRW background spacetime with a spatial curvature perturbation; In the next two, we added an evolving primordial tensor perturbation with just $h^{\times}$ or $h^{+}$ component to the spacetime and find that these two components leave a characteristic imprint on the distribution of the fractional density perturbation in the nonlinear region. More precisely: The primordial tensor modes do decay rapidly and never leave the linear regime as shown in Fig.~\ref{fig:h}. \textit{Before they die out, however, they have modified the profile of overdensity slightly, and the modification is amplified with time} as shown in the right one of Fig.~\ref{fig:overdense}. The Fig.~\ref{fig:xy1} shows the final modification in the nonlinear region, so the FIG.4 can only be produced by the nonlinear numerical relativity code which is a necessary here.

These imprints may be a possible probe of a GW background in the future. We do try to suggest exactly how or how much these tenor modes may affect observables including 2D angular power spectrum of large-scale structure tracers (as Fabian Schmidt \textit{et. al.} proposed \cite{Jeong:2012nu,Schmidt:2012nw}, but unfortunately here we just study the scalar modes with wave length $l=500$ and there is no suitable primordial power spectrum on hand.

And it's worth pointing out that much of the infrastructure for our work was developed and the development of tensor perturbations from scalar perturbations
was already demonstrated in the MacPherson \textit{et. al.} paper \cite{Macpherson:2016ict}. For simplicity and correctness, our work takes advantage of their configuration for simulation directly. That is to say their work serves as an important tool for our work. On the other hand our work also goes beyond their work: a) additional primordial tensor modes have been added to the initial conditions and their evolution has been shown in Fig.~\ref{fig:h}. b)their contribution to the evolution of overdensity has been shown in the right one of Fig.~\ref{fig:overdense}. c)their final imprint on the distribution of the fractional density perturbation in the nonlinear region also has been given in \ref{fig:xy1}.

\vspace{5mm}
\noindent {\bf Acknowledgments}
We  would  like  to  thank  Qing-Guo Huang  and  You-Jun Lu for their helpful discussions and advices on this paper. This work is partly supported by the National Natural Science Foundation of China under grant No. 11690024, the Strategic Priority Program of the Chinese Academy of Sciences (Grant No. XDB 23040100)

\begin{appendix}
\section{Conergence test}
\begin{figure}[]
\begin{center}
\includegraphics[scale=0.5]{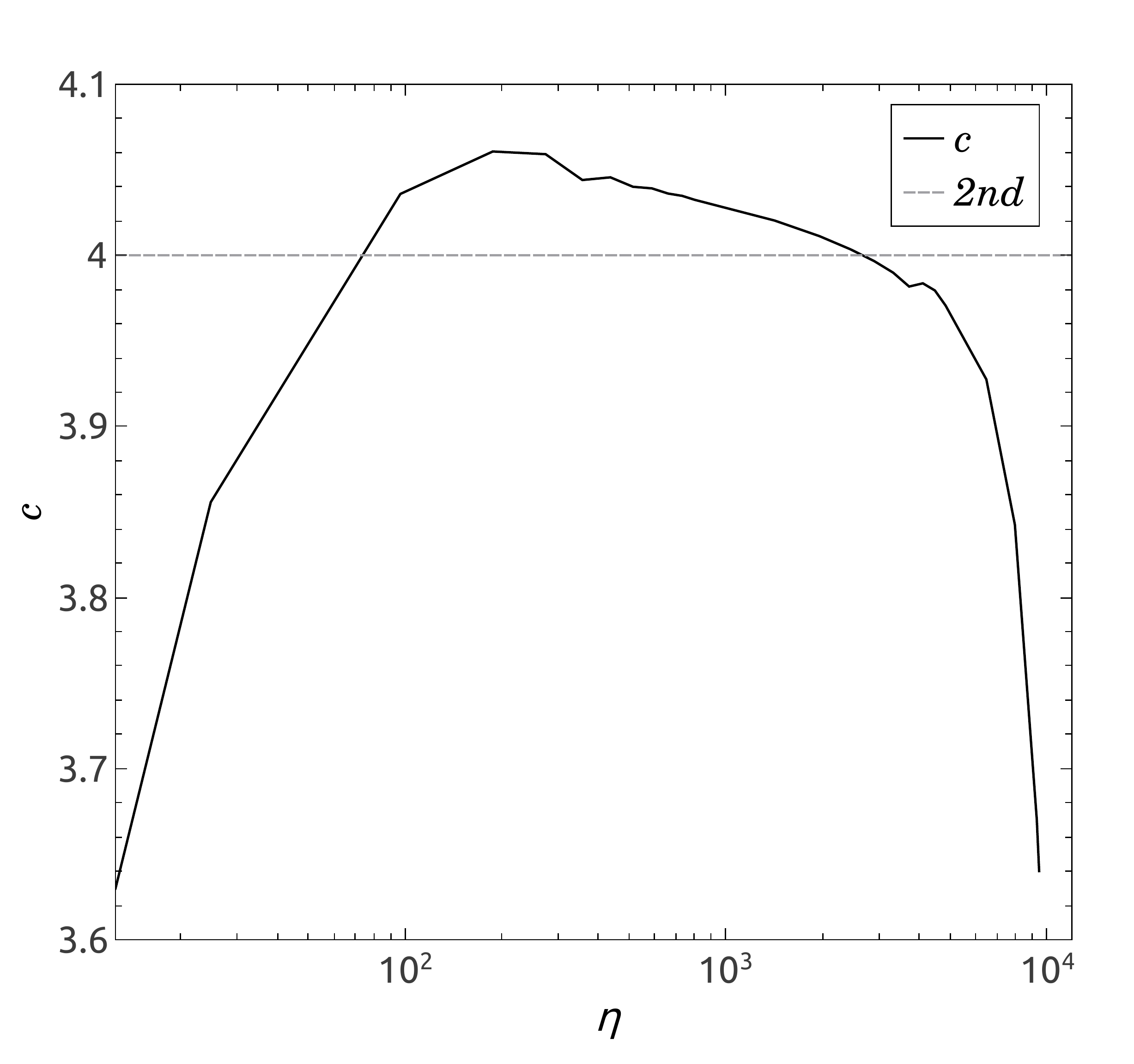}
\end{center}
\caption{Convergence factor from the three-point convergence test.}
\label{fig:c}
\end{figure}

Since the half wave length of the perturbations used in our paper is $250$, we perform the three-point convergence test under the following three resolutions: $(\frac{1000}{25})^3$, $(\frac{1000}{12.5})^3$ and $(\frac{1000}{6.25})^3$ where there are more enough grid points coinciding. Then the convergence factor of any output
\begin{equation}
c=\frac{\left\Vert O^{(coarse)}-O^{(mild)}\right\Vert}{\left\Vert O^{(mild)}-O^{(fine)}\right\Vert},
\end{equation}
will congverge to $2^n$ and $O$ will converge at $n$-th order. Fig.~\ref{fig:c} shows that the $c$ of $\rho_1+\rho_0$ remains close to the second-order value 4.

\begin{figure}[]
\begin{center}
\includegraphics[scale=0.3]{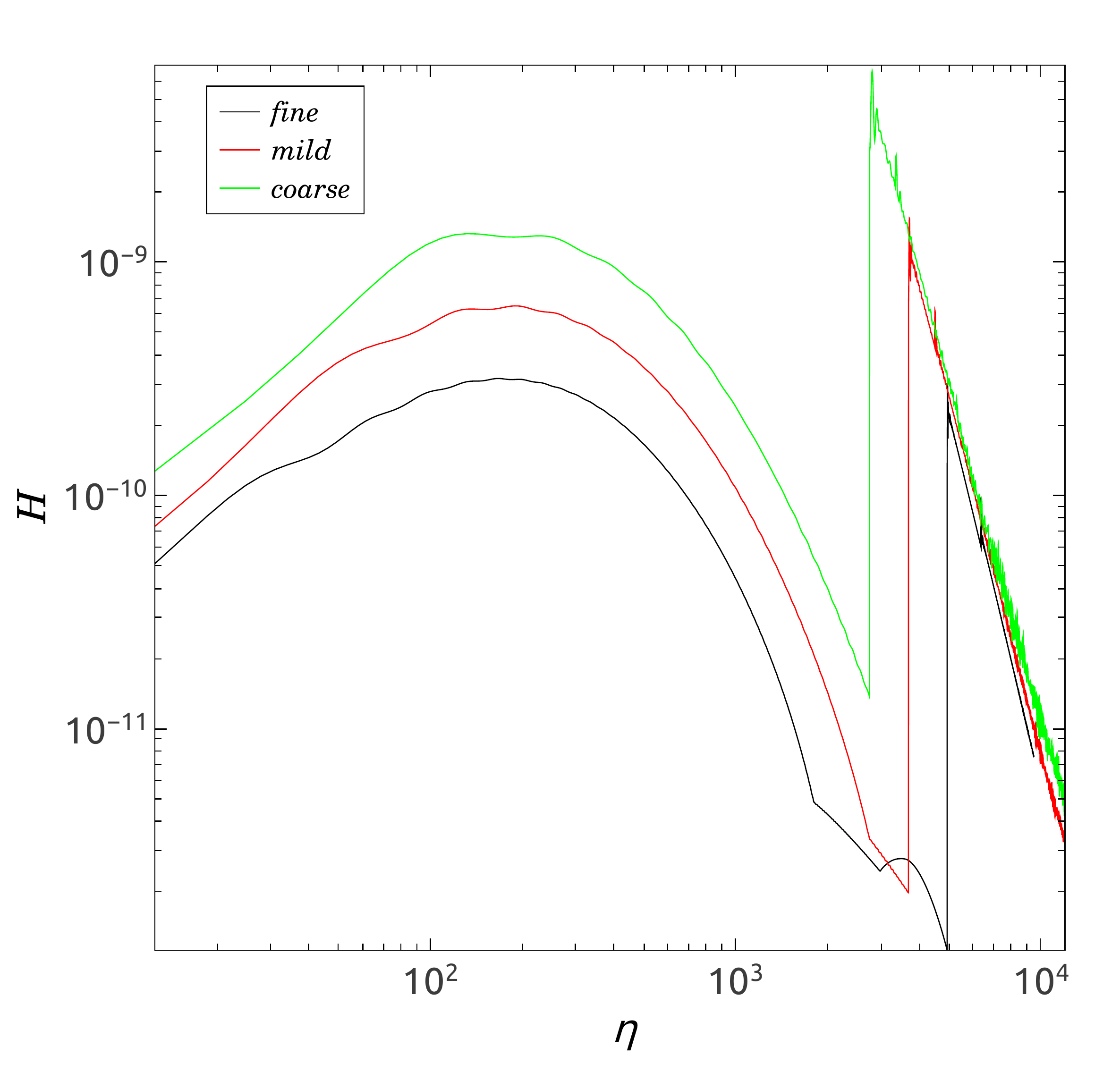}
\includegraphics[scale=0.3]{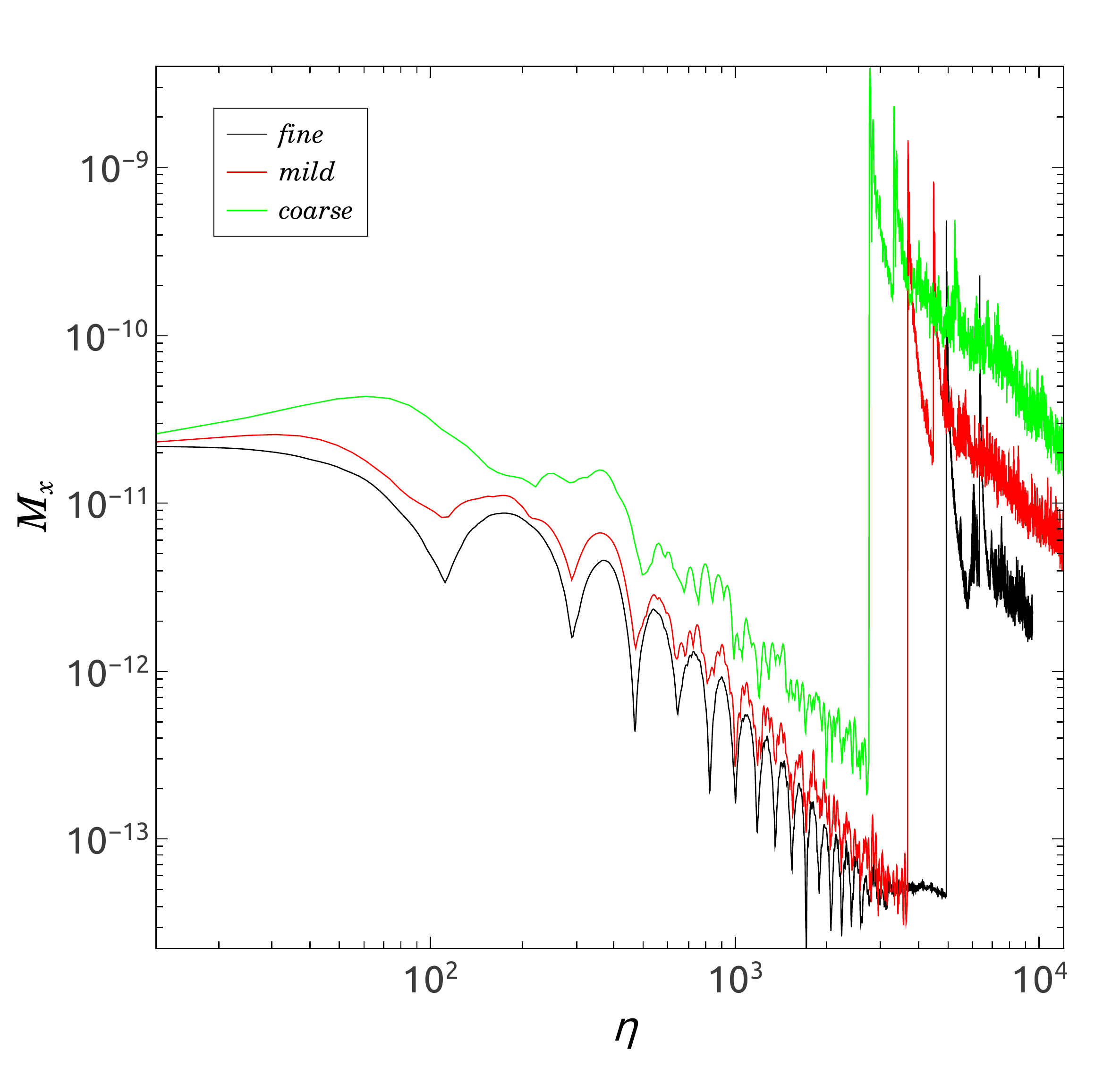}
\end{center}
\caption{Maximum of the Hamiltonian (left) and the x-component of momentum (right) constraint violation in simulations with $h^{\times}$.}
\label{fig:constraint}
\end{figure}

In our paper, we give the initial conditions by solving the perturbed Einstein equations with $\Phi_0=10^{-4}$. So there is a question that whether these initial data satisfy the Hamiltonian constraint and the momentum constraint or not. Given the 3-Riemann scalar $^{(3)}R$, the covariant derivative associated with the 3-metric $D_j$, and the matter energy and momentum density as measured by the Eulerian observer $E$ and $p_i$, we can spcify the form of the Hamiltonian constraint violation and the momentum constraint violation as
\begin{equation}
\mathcal{H}=\frac{1}{2}(^{(3)}R+K^2-K_{ij}K^{ij})-8\pi E
\end{equation}
and
\begin{equation}
\mathcal{M}_i=D_jK^j_i-D_iK-8\pi p_i.
\end{equation}
Fig.~\ref{fig:constraint} shows the evolution of the maximum of the Hamiltonian constraint violation and the x-component of momentum constraint violation under three resolutions: $(\frac{1000}{6.25})^3$, $(\frac{1000}{12.5})^3$ and $(\frac{1000}{25})^3$. We can see that the constraints remain bounded during the evolution and converge at first order with increasing resolution. In particular the strong initial increases and latter spikes shown in Fig.~\ref{fig:constraint} become gentle with increasing resolution.

\end{appendix}




\begin{thebibliography}{99}
\frenchspacing

\bibitem{Starobinsky:1980te}
  A.~A.~Starobinsky,
  Phys.\ Lett.\  {\bf 91B}, 99 (1980).
  doi:10.1016/0370-2693(80)90670-X

\bibitem{Guth:1980zm}
  A.~H.~Guth,
  Phys.\ Rev.\ D {\bf 23}, 347 (1981).
  doi:10.1103/PhysRevD.23.347

\bibitem{Seljak:1996gy}
  U.~Seljak and M.~Zaldarriaga,
  Phys.\ Rev.\ Lett.\  {\bf 78}, 2054 (1997)
  doi:10.1103/PhysRevLett.78.2054
  [astro-ph/9609169].

\bibitem{Kamionkowski:1996zd}
  M.~Kamionkowski, A.~Kosowsky and A.~Stebbins,
  Phys.\ Rev.\ Lett.\  {\bf 78}, 2058 (1997)
  doi:10.1103/PhysRevLett.78.2058
  [astro-ph/9609132].

\bibitem{Kamionkowski:2015yta}
  M.~Kamionkowski and E.~D.~Kovetz,
  Ann.\ Rev.\ Astron.\ Astrophys.\  {\bf 54}, 227 (2016)
  doi:10.1146/annurev-astro-081915-023433
  [arXiv:1510.06042 [astro-ph.CO]].

\bibitem{Book:2011dz}
  L.~Book, M.~Kamionkowski and F.~Schmidt,
  Phys.\ Rev.\ Lett.\  {\bf 108}, 211301 (2012)
  doi:10.1103/PhysRevLett.108.211301
  [arXiv:1112.0567 [astro-ph.CO]].

\bibitem{Masui:2010cz}
  K.~W.~Masui and U.~L.~Pen,
  Phys.\ Rev.\ Lett.\  {\bf 105}, 161302 (2010)
  doi:10.1103/PhysRevLett.105.161302
  [arXiv:1006.4181 [astro-ph.CO]].

\bibitem{Dodelson:2003bv}
  S.~Dodelson, E.~Rozo and A.~Stebbins,
  Phys.\ Rev.\ Lett.\  {\bf 91}, 021301 (2003)
  doi:10.1103/PhysRevLett.91.021301
  [astro-ph/0301177].

\bibitem{Dodelson:2010qu}
  S.~Dodelson,
  Phys.\ Rev.\ D {\bf 82}, 023522 (2010)
  doi:10.1103/PhysRevD.82.023522
  [arXiv:1001.5012 [astro-ph.CO]].

\bibitem{Jeong:2012nu}
  D.~Jeong and F.~Schmidt,
  Phys.\ Rev.\ D {\bf 86}, 083512 (2012)
  doi:10.1103/PhysRevD.86.083512
  [arXiv:1205.1512 [astro-ph.CO]].

\bibitem{Schmidt:2012nw}
  F.~Schmidt and D.~Jeong,
  Phys.\ Rev.\ D {\bf 86}, 083513 (2012)
  doi:10.1103/PhysRevD.86.083513
  [arXiv:1205.1514 [astro-ph.CO]].

\bibitem{Yadav:2010cc}
  J.~K.~Yadav, J.~S.~Bagla and N.~Khandai,
  Mon.\ Not.\ Roy.\ Astron.\ Soc.\  {\bf 405}, 2009 (2010)
  doi:10.1111/j.1365-2966.2010.16612.x
  [arXiv:1001.0617 [astro-ph.CO]].

\bibitem{Scrimgeour:2012wt}
  M.~Scrimgeour {\it et al.},
  Mon.\ Not.\ Roy.\ Astron.\ Soc.\  {\bf 425}, 116 (2012)
  doi:10.1111/j.1365-2966.2012.21402.x
  [arXiv:1205.6812 [astro-ph.CO]].

\bibitem{Amendola:2016saw}
  L.~Amendola {\it et al.},
  arXiv:1606.00180 [astro-ph.CO].

\bibitem{Maartens:2015mra}
  R.~Maartens {\it et al.} [SKA Cosmology SWG Collaboration],
  PoS AASKA {\bf 14}, 016 (2015)
  [arXiv:1501.04076 [astro-ph.CO]].

\bibitem{Ivezic:2008fe}
  Z.~Ivezic {\it et al.} [LSST Collaboration],
  arXiv:0805.2366 [astro-ph].

\bibitem{Buchert:1999er}
  T.~Buchert,
  Gen.\ Rel.\ Grav.\  {\bf 32}, 105 (2000)
  doi:10.1023/A:1001800617177
  [gr-qc/9906015].

\bibitem{Kolb:2004am}
  E.~W.~Kolb, S.~Matarrese, A.~Notari and A.~Riotto,
  Phys.\ Rev.\ D {\bf 71}, 023524 (2005)
  doi:10.1103/PhysRevD.71.023524
  [hep-ph/0409038].

\bibitem{Buchert:2007ik}
  T.~Buchert,
  Gen.\ Rel.\ Grav.\  {\bf 40}, 467 (2008)
  doi:10.1007/s10714-007-0554-8
  [arXiv:0707.2153 [gr-qc]].
\bibitem{Rasanen:2011ki}
  S.~Rasanen,
  Class.\ Quant.\ Grav.\  {\bf 28}, 164008 (2011)
  doi:10.1088/0264-9381/28/16/164008
  [arXiv:1102.0408 [astro-ph.CO]].

\bibitem{Clarkson:2011zq}
  C.~Clarkson, G.~Ellis, J.~Larena and O.~Umeh,
  Rept.\ Prog.\ Phys.\  {\bf 74}, 112901 (2011)
  doi:10.1088/0034-4885/74/11/112901
  [arXiv:1109.2314 [astro-ph.CO]].

\bibitem{Buchert:2011sx}
  T.~Buchert and S.~Rasanen,
  Ann.\ Rev.\ Nucl.\ Part.\ Sci.\  {\bf 62}, 57 (2012)
  doi:10.1146/annurev.nucl.012809.104435
  [arXiv:1112.5335 [astro-ph.CO]].

\bibitem{Buchert:2015iva}
  T.~Buchert {\it et al.},
  Class.\ Quant.\ Grav.\  {\bf 32}, 215021 (2015)
  doi:10.1088/0264-9381/32/21/215021
  [arXiv:1505.07800 [gr-qc]].

\bibitem{Green:2015bma}
  S.~R.~Green and R.~M.~Wald,
  arXiv:1506.06452 [gr-qc].

\bibitem{Buchert:2017obp}
  T.~Buchert,
  Mon.\ Not.\ Roy.\ Astron.\ Soc.\  {\bf 473}, no. 1, L46 (2018)
  doi:10.1093/mnrasl/slx160
  [arXiv:1704.00703 [astro-ph.CO]].

\bibitem{Adamek:2013wja}
  J.~Adamek, D.~Daverio, R.~Durrer and M.~Kunz,
  Phys.\ Rev.\ D {\bf 88} (2013) no.10,  103527
  doi:10.1103/PhysRevD.88.103527
  [arXiv:1308.6524 [astro-ph.CO]].

\bibitem{Giblin:2015vwq}
  J.~T.~Giblin, J.~B.~Mertens and G.~D.~Starkman,
  Phys.\ Rev.\ Lett.\  {\bf 116}, no. 25, 251301 (2016)
  doi:10.1103/PhysRevLett.116.251301
  [arXiv:1511.01105 [gr-qc]].

\bibitem{Mertens:2015ttp}
  J.~B.~Mertens, J.~T.~Giblin and G.~D.~Starkman,
  Phys.\ Rev.\ D {\bf 93}, no. 12, 124059 (2016)
  doi:10.1103/PhysRevD.93.124059
  [arXiv:1511.01106 [gr-qc]].

\bibitem{Bentivegna:2015flc}
  E.~Bentivegna and M.~Bruni,
  Phys.\ Rev.\ Lett.\  {\bf 116}, no. 25, 251302 (2016)
  doi:10.1103/PhysRevLett.116.251302
  [arXiv:1511.05124 [gr-qc]].

\bibitem{Loffler:2011ay}
  F.~Loffler {\it et al.},
  Class.\ Quant.\ Grav.\  {\bf 29}, 115001 (2012)
  doi:10.1088/0264-9381/29/11/115001
  [arXiv:1111.3344 [gr-qc]].

\bibitem{Macpherson:2016ict}
  H.~J.~Macpherson, P.~D.~Lasky and D.~J.~Price,
  Phys.\ Rev.\ D {\bf 95}, no. 6, 064028 (2017)
  doi:10.1103/PhysRevD.95.064028
  [arXiv:1611.05447 [astro-ph.CO]].

\bibitem{Brown:2008sb}
  J.~D.~Brown, P.~Diener, O.~Sarbach, E.~Schnetter and M.~Tiglio,
  Phys.\ Rev.\ D {\bf 79}, 044023 (2009)
  doi:10.1103/PhysRevD.79.044023
  [arXiv:0809.3533 [gr-qc]].

\bibitem{Reisswig:2010cd}
  C.~Reisswig, C.~D.~Ott, U.~Sperhake and E.~Schnetter,
  Phys.\ Rev.\ D {\bf 83}, 064008 (2011)
  doi:10.1103/PhysRevD.83.064008
  [arXiv:1012.0595 [gr-qc]].

\bibitem{code}
  McLachlan, a public BSSN code URL. http://www.cct.lsu.edu/~eschnett/    McLachlan/

\bibitem{Baumgarte:1998te}
  T.~W.~Baumgarte and S.~L.~Shapiro,
  Phys.\ Rev.\ D {\bf 59}, 024007 (1999)
  doi:10.1103/PhysRevD.59.024007
  [gr-qc/9810065].

\bibitem{Shibata:1995we}
  M.~Shibata and T.~Nakamura,
  Phys.\ Rev.\ D {\bf 52}, 5428 (1995).
  doi:10.1103/PhysRevD.52.5428

\bibitem{Alcubierre:2000xu}
  M.~Alcubierre {\it et al.},
  Phys.\ Rev.\ D {\bf 62}, 044034 (2000)
  doi:10.1103/PhysRevD.62.044034
  [gr-qc/0003071].

\bibitem{Moesta:2013dna}
  P.~Mösta {\it et al.},
  Class.\ Quant.\ Grav.\  {\bf 31}, 015005 (2014)
  doi:10.1088/0264-9381/31/1/015005
  [arXiv:1304.5544 [gr-qc]].

\bibitem{Baiotti:2004wn}
  L.~Baiotti, I.~Hawke, P.~J.~Montero, F.~Loffler, L.~Rezzolla, N.~Stergioulas, J.~A.~Font and E.~Seidel,
  Phys.\ Rev.\ D {\bf 71}, 024035 (2005)
  doi:10.1103/PhysRevD.71.024035
  [gr-qc/0403029].

\bibitem{Hawke:2005zw}
  I.~Hawke, F.~Loffler and A.~Nerozzi,
  Phys.\ Rev.\ D {\bf 71}, 104006 (2005)
  doi:10.1103/PhysRevD.71.104006
  [gr-qc/0501054].

\bibitem{Malik:2008im}
  K.~A.~Malik and D.~Wands,
  Phys.\ Rept.\  {\bf 475}, 1 (2009)
  doi:10.1016/j.physrep.2009.03.001
  [arXiv:0809.4944 [astro-ph]].

\bibitem{Wang:1995kb}
  Y.~Wang,
  Phys.\ Rev.\ D {\bf 53}, 639 (1996)
  doi:10.1103/PhysRevD.53.639
  [astro-ph/9501116].

\bibitem{Baumann:2007zm}
  D.~Baumann, P.~J.~Steinhardt, K.~Takahashi and K.~Ichiki,
  Phys.\ Rev.\ D {\bf 76}, 084019 (2007)
  doi:10.1103/PhysRevD.76.084019
  [hep-th/0703290].
\end{thebibliography}
\end{document}